\documentclass[aps,prl,twocolumn,showpacs,preprintnumbers,amsmath,amssymb]{revtex4}
\usepackage{amssymb}
\usepackage{amsmath}
\usepackage{graphicx}

\begin{document}

% Abbreviated title for the page headers
\title{Photoinduced dynamics in quantum rings }

% Authors
\author{Zhen-Gang Zhu and J. Berakdar}

\address{Institut f\"{u}r Physik, Martin Luther Universit\"{a}t
Halle-Wittenberg, Heinrich -Damerow-Stra{\ss}e, 4, 06120 Halle,
Germany }

\begin{abstract}
We investigate the spin-dependent dynamical response of a
semiconductor quantum ring with a spin orbit interaction (SOI) upon
the application of  a single and two linearly polarized, picosecond,
asymmetric electromagnetic
pulses in the presence of a static magnetic flux.
 We find that the pulse-generated electric dipole
moment (DM) is spin dependent. It is also shown that the
SOI induces an extra SU(2) effective flux in addition to the
static external magnetic flux which is reflected in an additional
periodicity of the spin-dependent  DM.
Furthermore, the pulses may induce a net dynamical charge
currents (CC) and  dynamical spin currents (SC) when
 the clockwise and anti-clockwise symmetry of the carrier is
broken upon the pulse application.
\end{abstract}

 \maketitle

\section{Introduction}
Spin-orbit interaction (SOI) in semiconducting low dimensional
structures is a key factor for spintronic research
\cite{spintronics}. There are two important kinds of SOI in
conventional semiconductors: one is the Dresselhaus SOI induced by
bulk inversion asymmetry \cite{dresselhaus}, and the other is the
Rashba SOI caused by structure inversion asymmetry \cite{rashba}. As
pointed out in \cite{lommer}, the Rashba SOI is dominant in a narrow
gap semiconductor and the strength of it can be tuned by an external
gate voltage \cite{materials}. This  tunability  of the magnitude
of the Rashba SOI is crucial for the operation of  spintronics device
such as the spin field effect transistor \cite{datta} and the spin
interference device \cite{nitta}.

In this work we are interested in quantum rings (QR) \cite{imry} which are synthesized
routinely with current nanotechnology.
Available  phase-coherent rings
vary in a wide range in size and particle density \cite{exp}.
On the theoretical side, the equilibrium properties of QRs
are fairly  understood and documented \cite{imry}. Current focus is on the
non-equilibrium dynamics, in particular that driven
 by external time-dependent
electromagnetic fields
\cite{pulseexp,matos}. E.g., it has been shown that the
irradiation with picosecond (from a few hundreds femtoseconds up to
nanoseconds, typically picsecond \cite{hcp}), time-asymmetric,
low-intensity light fields generates charge polarization and charge
currents in the ring.
 A particular feature of the driving pulse is that
 the electric field has  a short
 half cycle followed by a much longer and weaker half cycle of
an opposite polarity.  Such pulses are called half-cycle pulse (HCP)
because, under certain conditions,  only the very short and strong half optical cycle is
decisive for the carrier  dynamics.

Here we study  QRs as those fabricated out of a two
dimensional electron gas  formed between  heterojuctions of  III-V
and II-VI semiconductors. The influence of the SOI in QRs on the equilibrium
properties have  already been studied
\cite{zhu1,ringsoi}. In this work, we
shall consider the spin-dependent \emph{non-equilibrium} dynamic of
the ring with SOI driven by HCP's and in the presence of a magnetic
flux. We investigate two cases: applying single pulse and two
time-delayed pulses with non-collinear polarization axes.

\section{Theoretical Model}
\subsection{Hamiltonian}
%\begin{figure}[thb]%
%\includegraphics*[width=\linewidth,height=\linewidth]{g1}
%\caption{Schematic graphs are shown (a)the geometry, spin configuration and
%the applied pulses; (b) Two time-delayed HCPs. For the single pulse case, only $F_{1}$
%is applied. (c) Energy spectrum for a ring with SOI. $\Delta_{S}$ defines  the distance between the spectrum
%symmetry axis and the smallest nearest integer.} \label{fig1}
%\end{figure}
For  effective single particle
Hamiltonian of a one-dimensional (1D) ballistic QR with SOI we use
%\begin{equation}
$\hat{H}'=\hat{H}_{\mbox{SOI}}+\hat{H}_{1}(t)$ \cite{zhu1},
%\end{equation}
with
\begin{eqnarray}
\hat{H}_{\mbox{SOI}}&=&\frac{\mathbf{p}^{2}}{2m^{*}}+V(\mathbf{r})+\frac{\alpha_{R}}{\hbar}(\hat{\mathbf{\sigma}}\times\mathbf{p})_{z}
, \notag\\
\hat{H}_{1}(t)&=&-e\mathbf{r}\cdot\mathbf{E}(t)+\mu_{B}\mathbf{B}(t)\cdot\hat{\mathbf{\sigma}}.
\label{h1t}
\end{eqnarray}
where $\alpha_{R}$ is the SOI parameter, $V(\mathbf{r})$ is confinement potential, $\mathbf{E}(t)$ and $\mathbf{B}(t)$ are the electric
 and the magnetic fields   of the pulse. Integrating out the $r$ dependence
$\hat{H}_{\mbox{SOI}}$ reads in cylindrical coordinates
\cite{zhu1,ringsoi}
\begin{equation}
\hat{H}_{\mbox{SOI}}=\frac{\hbar\omega_{0}}{2}[(i\frac{\partial}{\partial\varphi}+\frac{\phi}{\phi_{0}}-\frac{\omega_{R}}{2\omega_{0}}\sigma_{r})^{2}-
(\frac{\omega_{R}}{2\omega_{0}})^{2}+\frac{\omega_{B}}{\omega_{0}}\sigma_{z}].
\label{hsoi1}
\end{equation}
$\phi_{0}=h/e$ is the flux unit, $\phi=B\pi a^{2}$ is the magnetic
flux, $a$ is the radius of the ring, $\sigma_{r}=\sigma_{x}\cos\varphi+\sigma_{y}\sin\varphi$,
$\hbar\omega_{0}=\hbar^{2}/(m^{*}a^{2})=2E_{0}$,
$\hbar\omega_{R}=2\alpha_{R}/a$, $\hbar\omega_{B}=2\mu_{B}B$ and an
external static magnetic field $\mathbf{B}=B\hat{\mathbf{e}}_{z}$.

The single-particle eigenstates of  $\hat{H}_{\mbox{SOI}}$ are
represented as
$\Psi_{n}^{S}(\varphi)=e^{i(n+1/2)\varphi}\nu^{S}(\gamma,\varphi)$
where $\nu^{S}(\gamma,\varphi)=( a^{S}e^{-i\varphi/2},
b^{S}e^{i\varphi/2})^{T}$ (\textit{T} means transposed) are spinors in the angle dependent local
frame,  and $a^{\uparrow}=\cos(\gamma/2),\;
b^{\uparrow}=\sin(\gamma/2),\; a^{\downarrow}=-\sin(\gamma/2),
b^{\downarrow}=\cos(\gamma/2),$ where
$\tan\gamma=-Q_{R}=-\omega_{R}/\omega_{0}$ (we ignore the Zeeman
splitting). $\gamma$ describes the direction of the spin
quantization axis. The energy spectrum of the QR with the SOI reads
$E_{n}^{S}=\frac{\hbar\omega_{0}}{2}\left[(n-\frac{\phi}{\phi_{0}}+\frac{1-Sw}{2})^{2}-\frac{Q_{R}^{2}}{4}\right]$,
where $w=\sqrt{1+Q_{R}^{2}}=1/\cos\gamma$, and $S=\pm1$ stands for
spin up (down) in the local frame.

\subsection{Time-dependent wave functions}
At first we apply a  single HCP pulse  at $t=0$. The pulse
propagates  in the $z$ direction and has a duration $\tau_{d}$.
Its E-field is along the $x$ axis. When two pulses are applied, the first
pulse is followed by a second one at $t=\tau$ with the same duration
but the E-field being along the $y$ axis.  We consider the
case where $\tau_d$ is much
shorter than the ballistic time of the QR carriers. The single particle
states develop then as \cite{matos}
\begin{eqnarray}
\Psi_{n}^{S}(\varphi,t>0)&=&\Psi_{n}^{S}(\varphi,t<0)e^{i\alpha_{1}\cos\varphi},
\notag \\
\Psi_{n_{0}}^{S_{0}}(\varphi,t>\tau)&=&\Psi_{n_{0}}^{S_{0}}(\varphi,t<\tau)e^{i\alpha_{2}\sin\varphi}, \notag \\
\alpha_{1(2)}=\frac{eap_{1(2)}}{\hbar}, &&
p_{1(2)}=-\int_{0}^{\tau_{d}}E_{1(2)}(t)dt,\notag\\
&&E_{1(2)}(t)=F_{1(2)}f(t).
\label{matching}
\end{eqnarray}
 $F_{1(2)}$ and $f(t)$ describe the
amplitude and the time dependence of the E-field of the first
(second) pulse respectively. The pulse effect is encapsulated
entirely in the \emph{action parameter} $\alpha_{1(2)}$. With the
initial conditions $n(t<0)=n_{0}$ and $S(t<0)=S_{0}$ one finds
\begin{equation}
\Psi_{n_{0}}^{S_{0}}(\varphi,t)=
\sum_{ns}\frac{C_{n}^{S}(n_{0}S_{0}t)}{\sqrt{2\pi}}e^{i(n+1/2)\varphi}e^{-iE_{n}^{S}t/\hbar}|\nu^{S}\rangle,
\label{wavef1}
\end{equation}
with
\begin{equation}
C_{n}^{S}=\left\{
\begin{array}{l l}
\delta_{SS_{0}}\delta_{nn_{0}} & \mbox{for t}\leq0,\\
\delta_{SS_{0}}i^{n_{0}-n}J_{n_{0}-n}(\alpha_{1}) & \mbox{for t}\in [0,\tau), \\
\sum_{n'}\Lambda^{SS_{0}}_{nn'n_{0}}
e^{i(E_{n}^{S}-E_{n'}^{S_{0}})\tau/\hbar} & \mbox{for t}>\tau,
\end{array}
\right. \label{coeff}
\end{equation}
where $\Lambda^{SS_{0}}_{nn'n_{0}}=\delta_{SS_{0}}[i^{n_{0}-n'}J_{n_{0}-n'}(\alpha_{1})J_{n-n'}(\alpha_{2})]$ and
$J_{n}$ is the n-th order Bessel function.

\section{Numerical Results and discussions}
\subsection{Single pulse case}
\begin{figure}[thb]%
\includegraphics*[width=\linewidth,height=\linewidth]{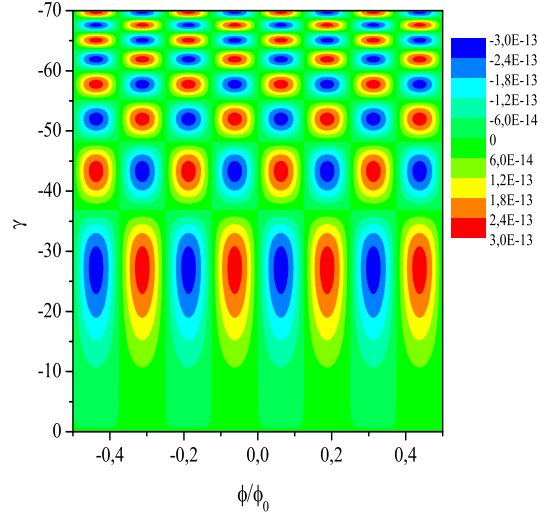}
\caption{Contour plot of the difference of the dipole moment for up
and down spins at the time moment $t/t_{p}=2$ ($t_p$ is the ballistic (field-free)
round trip time). The particle number is
 $N=100$ and the pulse properties are described by the dimensionless
 parameter
 $\alpha_1=0.1$ (cf. eq.(\ref{matching})).} \label{fig1}
\end{figure}
One HCP pulse induces dynamical oscillation of charge density which
is manifested as an electrical  dipole moment
$\mu_{n_{0}}^{S_{0}}(t)=ea\langle\cos\varphi\rangle_{n_{0}}^{S_{0}}(t)$,
where
$\langle\cos\varphi\rangle_{n_{0}}^{S_{0}}(t)=\int_{0}^{2\pi}d\varphi|
\Psi_{n_{0}}^{S_{0}}(\varphi,t)|^{2}\cos\varphi$. The total dipole
moment can be derived by summation over all the occupied states. In
the presence of the SOI, the dipole moment splits with respect to
different spin states. Fig. 1 shows a contour plot of the difference of
the dipole moments (in units of $ea$) for the up and down spins which varies  with
the magnetic flux and the SOI. The distinct positive and negative
regions correspond to the local spatial splitting of carrier density
for up and down spin states.
%Because the effect of the magnetic flux
%and the SOI enters into shift of the symmetry axis of the spectrum,
%the pattern shows oscillations with flux and the SOI.
The
oscillation with the static magnetic flux are observed, as
expected. The  oscillation with the SOI angle $\gamma$
are such,  larger $\gamma$ leads to shorter period
oscillations. Increasing $\gamma$ induces a shift of the
oscillation frequencies. However, increasing the strength of the
laser field brings more excited energy levels and more frequencies.

\subsection{Two pulses case}
\begin{figure}[thb]%
\includegraphics*[width=\linewidth,height=\linewidth]{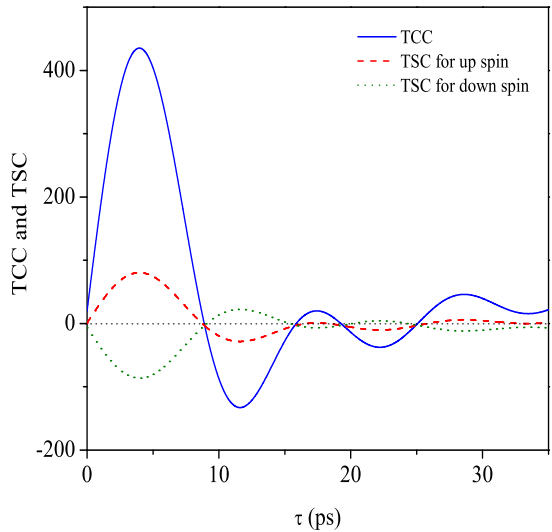}
\caption{TCC and TSC vary with the delay time $\tau$. The units for
CC and SC are $2E_{0}a/\phi_{0}$ and $E_{0}a/2\pi$ respectively. The
parameters are $N=100$, $a=400 nm$, $\gamma=-40^{\circ}$,
$F_{1}=F_{2}=500 V/cm$ and $\phi/\phi_{0}=0.3$.} \label{fig2}
\end{figure}
The charge current (CC) and the spin current (SC) are calculated
\cite{zhu2} using the velocity operator  $
\hat{\mathbf{v}}_{\varphi}=\hat{\mathbf{e}}_{\varphi}
\left\{\frac{-i\hbar}{m^{*}a}\partial_{\varphi}-\frac{\hbar}{m^{*}a}\frac{\phi}{\phi_{0}}
+\frac{\alpha_{R}}{\hbar}\sigma_{r}\right\}$. The SOI induces a
SU(2) vector potential (VP) appearing as the third term in the
velocity operator. The static magnetic flux and SU(2) VP generate
spin independent and spin-dependent persistent charge current (PCC)
respectively even in the absence of the pulse field. The laser pulse
triggers dynamic CC and SC which are  tunable by
external parameters. Therefore, the total CC (TCC) and the total SC (TSC)
(sum over the persistent and dynamic components) for up and down
spins are investigated in Fig. 2 with the delay time $\tau$.

It is clear, with the appropriate $\tau$, positive and negative TCC can
be obtained. TSC for up and down
spins are also oscillating and decaying with larger $\tau$. However
TSC shows opposite phase with respect to up and down spin states. If
$\phi=0$, the SOI gives rise to equal shifts for up and down spin
states but in opposite directions, making the TSC exactly reverse to
each other. If $\phi\neq 0$, the two components of TSC are not
exactly opposite, which corresponds to a slight imbalance occupation
for up and down spin states. More analysis of the behavior of the system driven by
the HCP pulse can be found in  Ref. \cite{zhu2}.

In summary, it is shown that asymmetric electromagnetic pulses
 can be used to generate and control spin-dependent charge
oscillation and dynamic currents in nano- and mesoscopic rings.

The work is support by the cluster of excellence "Nanostructured
Materials" of the state Saxony-Anhalt.

% Use the following code if you wish to generate your bibliography with BibTeX;
% replace the string "pss-demo" below with the name(s) of
% the BibTeX data base(s) you want to use.
% The resulting bibliography-output (the contents of the .bbl file)
% must be pasted back into this file before submission.
%
% \bibliographystyle{pss}
% \bibliography{pss-demo}
%
% Replace the following example bibliography with your references
% before submission:

\end{document}